\documentclass[%
      aps,
      prl, 
      reprint,
      twocolumn,
      superscriptaddress,
      showpacs]{revtex4-2}
      
\usepackage[english]{babel}
\usepackage{microtype}
\usepackage{graphicx}
\usepackage{booktabs}
\usepackage{lipsum}  
\usepackage[ colorlinks,
    linkcolor={blue},
    citecolor={blue},
    urlcolor={blue}]{hyperref}

\usepackage{tabularx}
\usepackage{multirow}
\usepackage{braket}
\usepackage{amsmath}
\usepackage{array}
\usepackage{setspace}
\usepackage{float}
\usepackage{ulem}
\usepackage{color}
\usepackage[varg]{txfonts}

\newcommand{\atgamma}{\bk{=}\bo}

\newcommand{\bo}{\mathbf{0}}

\newcommand{\bW}{\mathbf{W}}

\newcommand{\emikR}{e^{-i\mathbf{k}\cdot\mathbf{R}}}
\newcommand{\eikR}{e^{i\mathbf{k}\cdot\mathbf{R}}}
\newcommand{\bk}{\mathbf{k}}
\newcommand{\bq}{\mathbf{q}}
\newcommand{\bR}{\mathbf{R}}

\newcommand\eqt{\hspace{0.17em}{=}\hspace{0.17em}}

\newcommand\pt{\hspace{0.17em}{+}\hspace{0.17em}}
\newcommand\mt{\hspace{0.17em}{-}\hspace{0.17em}}
\newcommand\smalldist{\hspace{0.05em}}
\newcommand\kt{\hspace{0.17em}{<}\hspace{0.17em}}

  \newcommand\timest{\hspace{0.12em}{\times}\hspace{0.12em}}

\newcommand\br{\mathbf{r}}
\newcommand\bG{\mathbf{G}}

\definecolor{darkgreen}{rgb}{0.0,0.6,0.0}
\definecolor{orange}{rgb}{1.0,0.65,0.0}
\definecolor{brown}{rgb}{0.6,0.2,0.2}
\definecolor{gray}{rgb}{0.6,0.6,0.6}

\begin{document}

\title{Low-scaling \textit{GW} algorithm applied to twisted transition-metal dichalcogenide heterobilayers} 

\author{Maximilian Graml}
\affiliation{Institute of Theoretical Physics, University of Regensburg, 93053 Regensburg, Germany}
\affiliation{Regensburg Center for Ultrafast Nanoscopy (RUN), University of Regensburg, 93053 Regensburg, Germany}
  \author{Klaus Zollner}
  \affiliation{Institute of Theoretical Physics, University of Regensburg, 93053 Regensburg, Germany}
\author{Daniel \surname{Hernang\'{o}mez-P\'{e}rez}}
\affiliation{Department of Molecular Chemistry and Materials Science, Weizmann Institute of Science, Rehovot 7610001, Israel}
\author{Paulo E. Faria Junior}
\affiliation{Institute of Theoretical Physics, University of Regensburg, 93053 Regensburg, Germany}
\author{Jan Wilhelm}\email{jan.wilhelm@physik.uni-regensburg.de}
\affiliation{Institute of Theoretical Physics, University of Regensburg, 93053 Regensburg, Germany}
\affiliation{Regensburg Center for Ultrafast Nanoscopy (RUN), University of Regensburg, 93053 Regensburg, Germany}

 \linespread{1.1}
 \fontsize{10}{12}\selectfont
 \normalem

\begin{abstract}
The \textit{GW} method is widely used  for calculating the electronic band structure of materials. 
The high computational cost of \textit{GW} algorithms prohibits their application to many systems of interest. 
We present a periodic, low-scaling and highly efficient \textit{GW} algorithm that benefits from the locality of the  Gaussian basis and the polarizability.
The algorithm enables $G_0W_0$ calculations on a MoSe$_2$/WS$_2$ bilayer with 984 atoms per unit cell, in  42 hours using 1536 cores. 
This is four orders of magnitude faster than a plane-wave  $G_0W_0$ algorithm, allowing for unprecedented computational studies of electronic excitations at the nanoscale. 

\end{abstract}

\maketitle

\begin{figure}[b]
\centering
\includegraphics[width=\columnwidth]{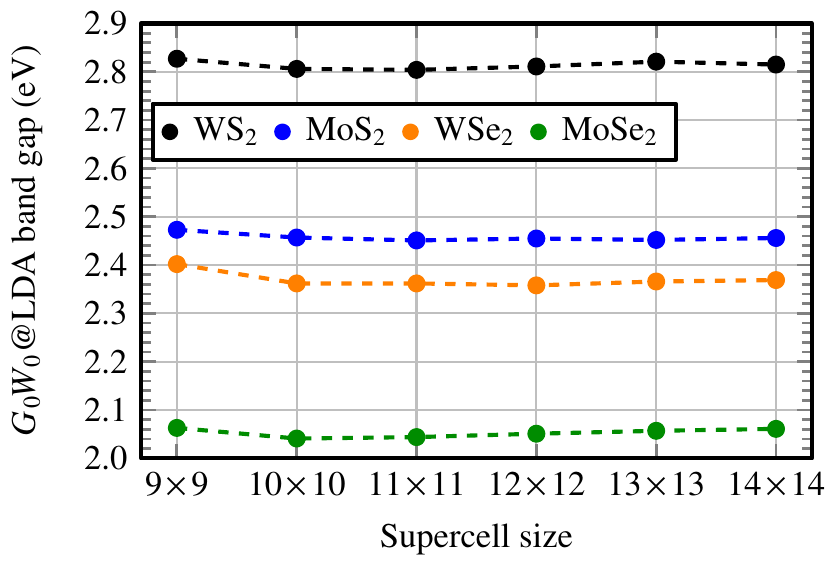}
\caption{
$G_0W_0$ bandgap  of monolayer WS$_2$, MoS$_2$, WSe$_2$ and MoSe$_2$ calculated from Eq.~\eqref{e18} as function of the supercell size (TZVP-MOLOPT basis set~\cite{Vandevondele2007}, without spin-orbit coupling (SOC)).
}
\label{f1}
\end{figure}

Electronic excitations in matter play a pivotal role in various physical phenomena, including light absorption and transport. 
The characteristics of these excitations are strongly influenced by the host material. 
Excitons, which are bound electron-hole pairs, exhibit a remarkable and unusually strong electron-hole binding in low-dimensional semiconductors that have emerged in the last decade~\cite{Mak2010}.
When stacking two atomically thin semiconductors on top of each other, the atomic alignment between the layers can exhibit  periodic variations, leading to a new type of in-plane superlattice known as the moiré superlattice.
Excitons in moir\'e structures have gained enormous attention recently~\cite{Seyler2019,Zhang2020:NC, Shabani2021, Karni2022, Schmitt2022, Gobato2022:NL,Barre2022:S,Naik2022,Rivera2018,Jin2018,Huang2022:NN} thanks to  their highly unusual exciton properties which 
include spatial confinement due to the moir\'e potential~\cite{Seyler2019}, interlayer~\cite{Schmitt2022,Karni2022}, and intralayer charge transfer~\cite{Naik2022}.
Furthermore, electronic properties of moir\'e lattices can be tuned by the band alignment and the twist angle between the layers such that moir\'e structures hold great promise as an exciting platform for probing electronic and photonic quantum phenomena over the next decade~\cite{Huang2022:NN}.
Gaining insights into excitons in moiré structures can be achieved through a combination of experiments, theoretical models, and computations. 
As an example, low-angle MoSe$_2$/WS$_2$ moir\'e structures have shown an interesting interplay of intra- and interlayer exciton hybridization because of the nearly degenerate conduction bands of the MoSe$_2$ and WS$_2$ layers. 
The conduction band offset and the wavefunction hybridization between  layers, however, is still under debate~\cite{Alexeev2019Nature,Tang2020NatNano,Zhang2020:NC,Ma2021NL,Gobato2022:NL}. 
Detailed knowledge about the electronic band structure of the MoSe$_2$/WS$_2$ moir\'e bilayer and the implication on exciton formation and binding is thus crucial to resolve this controversy.
In this work, we focus on the $GW$ method from many-body-perturbation theory~\cite{Hedin1965,Onida2002,Golze2019} which is an approximation for the electronic self-energy that allows for computing the electronic band structure of a given material.
Importantly, $GW$ accounts for the nonlocal, frequency-dependent screening of the interaction between electrons which is crucial in moir\'e bilayers. 
The $GW$ band structure is then the basis for the description of excitons via the Bethe–Salpeter equation~\cite{Onida2002,Blase2020}. 
Currently available plane-wave-based $GW$ algorithms are however incapable of treating low-angle moir\'e cells that contain thousands of atoms~\cite{Li2023}, despite their computational scalability to the largest supercomputers~\cite{Sangalli2019,DelBen2019,DelBen2020,Yu2022,Kundu2022}. 
Stochastic \textit{GW} methods may enable large-scale \textit{GW} calculations~\cite{Neuhauser2014,Brooks2020}, but it is not clear whether the numerical precision of this approach is sufficient for its application across the whole chemical space~\cite{Vlcek2017}. 
For computing the $GW$ band structure in large moiré cells, pristine unit-cell matrix projection (PUMP) has been suggested~\cite{Naik2022,Li2023}. 
PUMP is based on expanding the moiré cell wavefunctions in terms of the pristine unit-cell wavefunctions.
By construction, PUMP cannot capture nanometer-scale atomic reconstruction of moir\'e structures which can dramatically influence their electronic band structure~\cite{Shabani2021}.
The $GW$ space-time method~\cite{Rojas1995} offers a  promising route towards large-scale $GW$ calculations.
This is because the computational scaling is reduced from $O(N^4_\text{at}N_k^2)$ for standard $GW$ algorithms to $O(N^3_\text{at}N_k)$ in the \textit{GW} space-time method, where $N_\text{at}$ is the number of atoms in the unit cell and $N_k$ the number $k$-points used to discretize the Brillouin zone.
For achieving the scaling reduction, it is required to use a spatially local basis instead of plane waves. 
The local basis can be chosen as real-space grid where studies of unit cells up to one hundred atoms have been reported~\cite{Rojas1995,Liu2016}. 
Another choice of the spatially local basis is an atomic-orbital-like basis~\cite{Rohlfing1995}.
This choice is highly efficient in the $GW$ space-time method enabling $GW$ calculations on molecules with more than 1000 atoms~\cite{Wilhelm2018,Foerster2020,Duchemin2021,Wilhelm2021,Foerster2022}.
Periodic boundary conditions in the $GW$ space-time method with atomic-orbital-like basis functions have not been reported yet.
The main inhibiting factor has been the inclusion of $k$-dependent Coulomb interactions which represent a major challenge regarding computational efficiency and numerical precision~\cite{Ren2021,Zhu2021,Qiu2016}.
In this work, we overcome this challenge by employing real space representations of the polarizability, the screened Coulomb interaction and the self-energy. 
The real-space representation allows us to use the minimum image convention (MIC)~\cite{Paier2009,Irmler2018}, i.e., each atomic orbital in the simulation interacts only with the closest image of another atomic orbital.
We benchmark the algorithm on $G_0W_0$ bandgaps of monolayer~MoS$_2$, MoSe$_2$, WS$_2$, and WSe$_2$ finding an average deviation of only 0.06 eV from reference calculations~\cite{Gjerding2021,Camarasa2023}.
We also apply the $GW$ algorithm to a MoSe$_2$/WS$_2$ bilayer with an unprecedented cell size of 984 atoms which has an order of magnitude more atoms than previous state-of-the-art large-scale \textit{GW} calculations~\cite{Kundu2022}. 
We start with details on the main algorithmic advances for achieving large-scale \textit{GW} calculations on two-dimensional semiconductors. 
The full $GW$ algorithm is given in the Supporting Information. 
Following our previous work~\cite{Wilhelm2021,Wilhelm2016}, we compute the irreducible polarizability $\chi_{PQ}(\atgamma,i\tau)$ in imaginary time $i\tau$ at the $\Gamma$-point in an auxiliary atomic-orbital-like Gaussian basis set with indices $P,Q$.
The polarizability $\chi_{PQ}(\bk,i\tau)$ is however needed on a dense $k$-point mesh because it is later multiplied with the bare Coulomb interaction that diverges at the $\Gamma$-point and thus requires a fine $k$-point sampling. 
The atom-centered basis allows us to decompose the $\Gamma$-point result, $\chi_{PQ}(\atgamma,i\tau)$, using the identity
\begin{align}
\chi_{PQ}(\atgamma,i\tau)\eqt \sum_\mathbf{R}\chi_{PQ}^\mathbf{R}(i\tau)\,,\;\;\chi_{PQ}^\mathbf{R} \eqt \braket{\varphi^\mathbf{0}_P|\chi|\varphi^\mathbf{R}_Q}\,,\label{e6}
\end{align}
where $\chi_{PQ}^\mathbf{R}$ is the real-space representation of the polarizability and $\varphi_P^\bR$ denotes a Gaussian which is localized in  cell~$\bR$. 
For non-metallic systems, the polarizability $\chi(\mathbf{r},\mathbf{r}',i\tau)$ is \textit{space-local}, i.e.~$\chi(\mathbf{r},\mathbf{r}',i\tau)$ exponentially decays with  increasing~$|\mathbf{r}\mt\mathbf{r}'|$.~\cite{Kohn1996,Martin2016}
The matrix element $\chi_{PQ}^\mathbf{R}$ thus  vanishes in case of a large  distance between the center of~$\varphi_P^\mathbf{0}$ and the center of~$\varphi_Q^\mathbf{R}$.
We employ MIC, i.e., we assume that $\chi_{PQ}^\mathbf{R}(i\tau)$ in Eq.~\eqref{e6}  is non-zero only if the atomic center of~$\varphi_P^\mathbf{0}$ and  the atomic center of~$\varphi_Q^\mathbf{R}$ are closest together among all cells~$\bR$.
In this way, we extract $\chi_{PQ}^\mathbf{R}(i\tau)$ from  Eq.~\eqref{e6},
\begin{align}
\chi_{PQ}^\mathbf{R}(i\tau) = \left\{  
\begin{array}{cl} 
\chi_{PQ}(\atgamma,i\tau) & \text{\;\;if $\varphi_P^\mathbf{0}$, $\varphi_Q^\mathbf{R}$ closest\,,} \\[0.5em]
0 & \text{\;\;else\,,}\end{array}
\right.
\label{e7}
\end{align}
which is exact in the limit of a large, non-metallic unit cell. 
Using Eq.~\eqref{e7}, we obtain the polarizability at any $k$-point at negligible computational cost,
\begin{align}
\chi_{PQ}(\mathbf{k},i\tau) = \sum_{\mathbf{R}}\emikR\, \chi_{PQ}^\mathbf{R}(i\tau)\,.\label{e8}
\end{align}
Following the \textit{GW} space-time method~\cite{Rojas1995}, we compute the screened interaction in real space (full algorithm in SI),
\begin{align}
W_{PQ}^\mathbf{R} (i\tau) \coloneqq\braket{\varphi^\mathbf{0}_P|W(i\tau)|\varphi^\mathbf{R}_Q} \,,
\label{e14}
\end{align}
leading to the self-energy $\Sigma(\mathbf{r},\mathbf{r}',i\tau)\eqt iG(\mathbf{r},\mathbf{r}',i\tau)W(\mathbf{r},\mathbf{r}',i\tau)$ \cite{Rojas1995}.
$\Sigma(\mathbf{r},\mathbf{r}',i\tau)$ is space-local as~$G(\mathbf{r},\mathbf{r}',i\tau)$ is space-local~\cite{Kohn1996} and only elements of $W(\mathbf{r},\mathbf{r}',i\tau)$ with small~$|\mathbf{r}\mt\mathbf{r}'|$ contribute to $\Sigma$.
We thus continue with the minimum image of Eq.~\eqref{e14}
\begin{align}
W_{PQ}^\text{MIC} (i\tau) \coloneqq W_{PQ}^{\mathbf{R}_{PQ}^\text{min}}(i\tau)\label{e15}\,,
\end{align}
where the cell vector
\begin{align}
\mathbf{R}_{PQ}^\text{min}\eqt \underset{\bR}{\text{argmin}} \left|\bR_P - (\bR_Q + \bR)\right|
\end{align}
gives the smallest distance between the atomic centers $\bR_P$ of~$\varphi_P^\mathbf{0}$ and the atomic center $\bR_Q \pt\bR$ of ~$\varphi_Q^{\mathbf{R}}$.
We use $\bW^\text{MIC} (i\tau)$ to calculate the self-energy~$ \Sigma_{\mu\nu}(\atgamma,i\tau )$ in the atomic-orbital basis $\{\phi_\mu(\br)\}$ at the $\Gamma$-point.
We thus  avoid $k$-point sampling in this computationally expensive step. 
$k$-points in $\Sigma$ follow from MIC at negligible computational cost, cf.~Eqs.~\eqref{e7}, \eqref{e8},
\begin{align}
\Sigma_{\mu\nu}(\bk,i\tau) &
=\hspace{-0.3em}
\sum_\mathbf{R} \eikR\cdot
\left\{  
\begin{array}{cl} 
\Sigma_{\mu\nu}(\atgamma,i\tau) & \text{if $\phi_\mu^\mathbf{0}$, $\phi_\nu^\mathbf{R}$ closest\,,} \\[0.5em]
0 & \text{else\,.}\end{array}
\right.
\label{e17}
\end{align}
We transform the self-energy to real energy~\cite{Golze2019} and the Bloch basis  which allows us to compute  quasiparticle energies~$\varepsilon_{n\bk}^{G_0W_0}$,
\begin{align}
\varepsilon_{n\bk}^{G_0W_0} = \varepsilon_{n\bk} + \text{Re}\,\Sigma_{n\bk}(\varepsilon_{n\bk}^{G_0W_0}) -v^\text{xc}_{n\bk}\,,
\label{e18}
\end{align}
where $v^\text{xc}_{n\bk}$ is the diagonal of the exchange-correlation matrix.
The numerical trick in the presented $GW$ algorithm is the MIC used in Eqs.~\eqref{e7},~\eqref{e15}, and~\eqref{e17}.
MIC is exact in the limit of a large unit cell.
We determine the critical cell size for the validity of MIC by computing the $G_0W_0$ bandgap of monolayer MoS$_2$, MoSe$_2$, WS$_2$, and WSe$_2$, presented in Fig.~\ref{f1}.
For the four materials, the bandgap changes on average by only 11\,meV between the 10\,$\times$\smalldist10 supercell (300 atoms in the unit cell) and the 14\,$\times$\smalldist14 supercell (588 atoms in the unit cell).
We conclude that the $GW$ algorithm from this work can be used to study unit cells which are as large as a 10\,$\times$\smalldist10 supercell or larger. 
In the  Supporting Information, we show additional convergence tests on the basis set size, the number of time and frequency points, the $k$-point mesh size, filter threshold for sparse operations, and the vertical box height.

\begin{figure}[tb]
\centering
\includegraphics[width=\columnwidth]{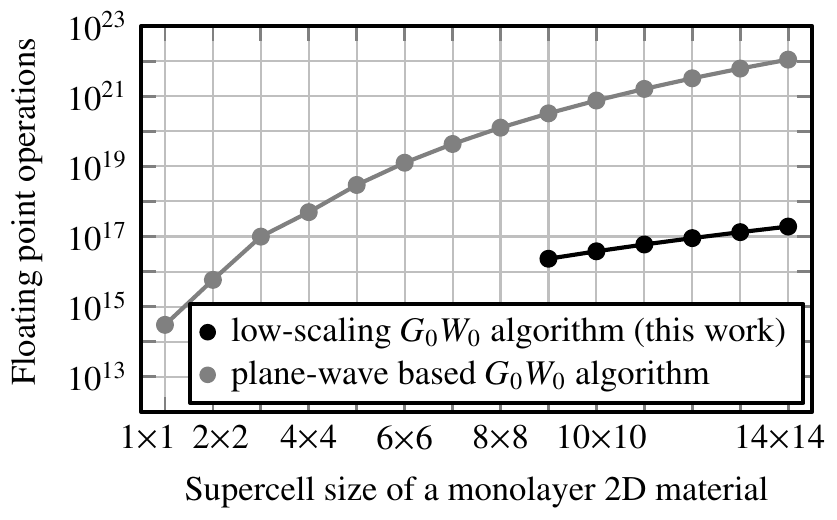}
\caption{
Number of floating point operations (real double precision) needed for executing $G_0W_0$ algorithms.
Black: low-scaling $G_0W_0$ algorithm from this work using a TZVP-MOLOPT basis set~\cite{Vandevondele2007}, gray: plane-wave-based $G_0W_0$ algorithm, Eq.~\eqref{e9}.
Underlying computational parameters are typical for mono\-layer MoS$_2$, MoSe$_2$, WS$_2$ and WSe$_2$, see detailed raw data available in the SI. 
}
\label{f2a}
\end{figure}

\begin{table}
 \caption{ $G_0W_0$@PBE bandgap (in eV, without SOC) of monolayer WS$_2$, MoS$_2$, WSe$_2$ and MoSe$_2$ computed from Eq.~\eqref{e18} (TZV2P-MOLOPT basis~\cite{Vandevondele2007}, 10\,$\times$\,10 supercell, detailed convergence test in the SI) and computed from plane-wave codes~\cite{Gjerding2021,Qiu2016,Camarasa2023}.
 }
 \begin{ruledtabular}
    \begin{tabular}{lcccc}
    Software package & MoS$_2$ & MoSe$_2$ & WS$_2$ & WSe$_2$ 
    \\
    \hline
This work, (CP2K~\cite{Kuehne2020,CP2K})  & 2.47  & 2.07& 2.81  &  2.37 
            \\
 GPAW~\cite{Gjerding2021} (SOC removed)  & 2.53 & 2.12 & 2.75 & 2.30 
        \\
BerkeleyGW~\cite{Qiu2016} (details in SI)  & 2.45 & 2.09 & 2.61 & 2.34 
      \\
 VASP~\cite{Camarasa2023}  & 2.50 & 2.06 & 2.70 & 2.34 
    \end{tabular}
 \end{ruledtabular}
     \label{t1}
 \end{table}
We compare the $G_0W_0$ bandgap of monolayer MoS$_2$, MoSe$_2$, WS$_2$, and WSe$_2$  to the $G_0W_0$ bandgap computed from three different plane-wave codes~\cite{Gjerding2021,Qiu2016,Camarasa2023}, see Table~\ref{t1}.
We find that our $G_0W_0$ bandgaps deviate on average by only 0.06 eV to the bandgaps from plane-wave based codes.
This small discrepancy might be due to the use of different pseudopotentials and the difficulty to reach the complete-basis-set limit. 
The presented algorithm has several computational advantages over  plane-wave-based $GW$ algorithms. 
The computational bottleneck in plane-wave-based $GW$ algorithms is the calculation of the irreducible polarizability~\cite{DelBen2019,Sangalli2019,Bruneval2008},
\begin{align}
 \begin{split}
  \chi_{\bG\bG'}&(\bq,i\omega) = 
  \sum_n^\text{occ} \sum_{n'}^\text{empty} \sum_\bk
   \frac{1}{\varepsilon_{n\bk+\bq}-\varepsilon_{n'\bk}+i\omega} 
  \\[0.5em] & \times \braket{n\bk{+}\bq|e^{i(\bq{+}\bG){\cdot}\br}|n'\bk}\braket{n'\bk|e^{-i(\bq{+}\bG'){\cdot}\br}|n\bk{+}\bq}\,,
  \end{split}
  \label{e9}
\end{align}
where $\bG,\bG'$ are reciprocal lattice vectors characterizing the plane wave $e^{i\bG{\cdot}\br}$, $\bq$ is a vector in the first Brillouin zone, $n, n'$ refer to occupied and empty bands, respectively, and the brackets in the second line denote integrals of a plane wave and Bloch states. 
The matrix in Eq.~\eqref{e9} is evaluated up to $|G^2| \kt |E_\text{cut}|$ for both $\bG$ and $\bG'$ where $E_\text{cut}$ is the dielectric energy cutoff.
We calculate the number of floating point operations necessary to perform the multiplications in Eq.~\eqref{e9}, see gray traces in Fig.~\ref{f2a}.
The estimate corresponds to the computational effort of a plane-wave based $G_0W_0$ algorithm for 2D semiconductors and is based on realistic numerical parameters used in large-scale $G_0W_0$ calculations~\cite{Qiu2016,Kundu2022}.
We also report the required  number of operations of our presented $G_0W_0$ algorithm in Fig.~\ref{f2a} (black traces).
Our $G_0W_0$ algorithm requires a similar number of floating point operations for a $9\times 9$ supercell as a plane-wave $G_0W_0$ algorithm for a $2\timest2$ supercell.
For a $14\timest14$ supercell, our $G_0W_0$ algorithm requires  $10^5$ times less operations  compared to a plane-wave based algorithm. 
This large factor has several origins, most important are the following: 
The plane-wave basis~$\{e^{i\bG{\cdot}\br}\}$ resolves large vacuum regions~\cite{Qiu2016,Kundu2022} for two-dimensional materials and is thus a factor 10 larger than the Gaussian auxiliary basis~$\{\varphi_P\}$.
We thus need to calculate $100$ times less matrix elements of $\chi$ in a Gaussian basis compared to Eq.~\eqref{e9}. 
Integrals over Gaussians, similar to the second line of Eq.~\eqref{e9}, are sparse due to the spatial locality of Gaussians~\cite{Wilhelm2021}. 
Only 3\,\% of the integrals need to be considered for a $14\timest14$ supercell reducing the number of operations by another factor 30.
In the present algorithm, $\chi$ is evaluated at the $\Gamma$-point using real-valued matrix algebra~\cite{Wilhelm2021} which makes another factor 4 compared to the complex matrix algebra in Eq.~\eqref{e9}.
In Eq.~\eqref{e9}, at least a $3\timest3$ mesh for $\bq$ is necessary~\cite{Kundu2022} which is responsible for another factor of~5~\footnote{Considering time-reversal symmetry, only 5 irreducible $k$-points are contained in a $3\timest3$ mesh.}.
These numerical parameters thus explain a factor of 60.000 between the required operations of a plane-wave $G_0W_0$ algorithm and the $G_0W_0$ algorithm from this work. 

Further advantages compared to plane-wave based algorithms include the cheap diagonalization of the Kohn-Sham matrix to obtain Bloch states thanks to the small Gaussian basis.
Also,  non-periodic directions are easily dealt with in our $GW$ algorithm by restricting the  sum  over  cells~$\bR$ to  periodic directions. 
It is not necessary to truncate the Coulomb operator~\cite{Qiu2016,Hueser2013} in non-periodic directions as in plane-wave algorithms.
Moreover, the self-energy~\eqref{e17} is available in the Gaussian basis set which allows to compute the $G_0W_0$ correction for all Bloch states at low computational cost. 

\begin{figure}[tb]
\centering
\includegraphics[width=\columnwidth]{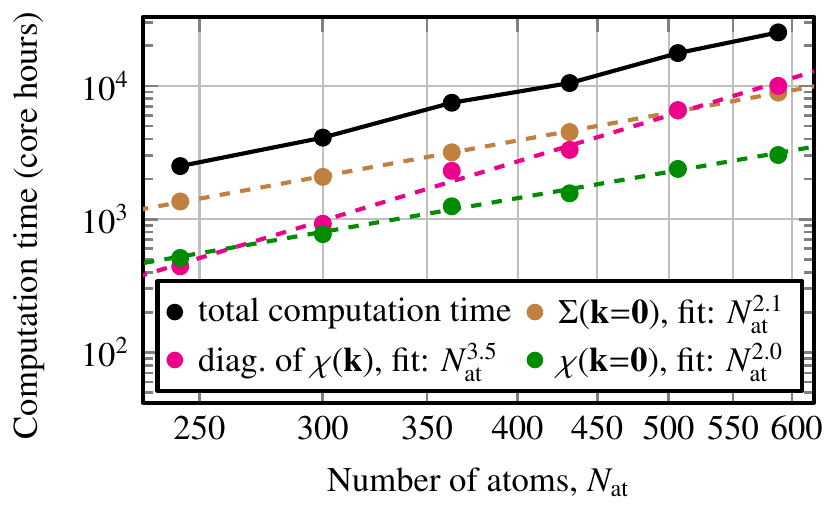}
\caption{
Execution time of a $G_0W_0$ calculation for MoSe$_2$ $9\timest9$\,--\,$14\timest14$ supercells (TZVP-MOLOPT basis set) on Supermuc-NG (Intel Skylake Xeon Platinum 8174).
Magenta points show the computational cost to diagonalize the polarizability~$\chi(\bk)$ which allows us to remove all spurious negative eigenvalues of $\chi(\bk)$ to ensure numerical stability. 
Dashed lines show a fit~$\alpha N_\text{at}^\beta$ to the execution time, where $\alpha$ and $\beta$ are fit parameters. 
Raw data is available in the SI. 
}
\label{f3}
\end{figure}

We measure the computation time of the algorithm, shown in Fig.~\ref{f3}. 
The computation time is moderate; as an example, a $G_0W_0$ calculation on the $10\timest10$ MoSe$_2$ supercell (300 atoms) takes only 7 hours on 576 cores. 
 Assuming ideal scalability starting from the $9\timest9$ cell, we estimate that a $G_0W_0$ calculation on 4500 atoms is in reach~\footnote{The computation time is \begin{align}
T = 
\left( \frac{N_\text{at}}{243} \right)^2
\,1870\;\text{core hours} +
\left( \frac{N_\text{at}}{243} \right)^3
\,443\;\text{core hours} \label{eq-footnote}
\end{align}
where we took the execution time of quadratic and cubic steps as  shown in Fig.~\ref{f3}. The maximum job size on Supermuc-NG is 150\,000 cores for 24 hours, making 3.6 Mio.~core hours, which will allow for a $G_0W_0$ calculation on 4500 atoms [Eq.~\eqref{eq-footnote}]. With restarting, even larger $G_0W_0$ calculations will become possible.}. 
Scalability improvements are subject of ongoing work to achieve this system size in practice. 

\begin{figure}[tb]
\centering
\includegraphics[width=1.0\columnwidth]{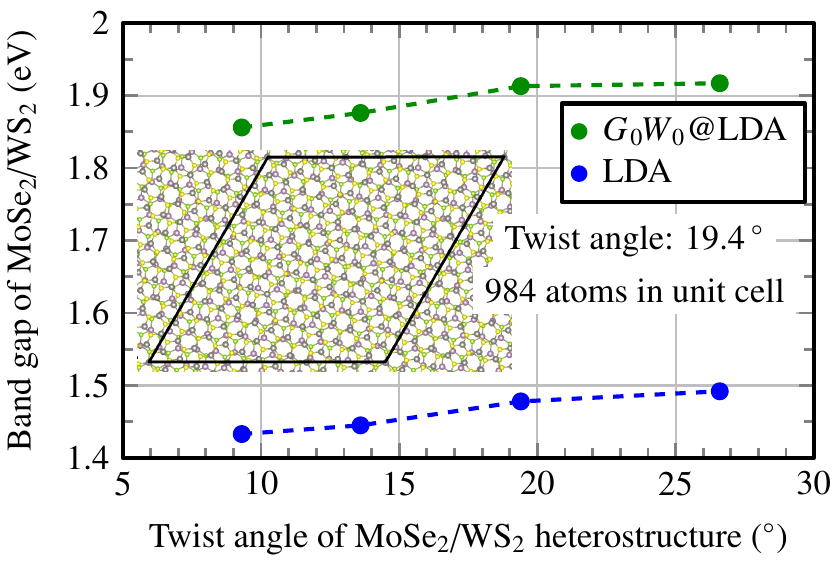}
\caption{
Band gap of a MoSe$_2$/WS$_2$ heterostructure as function of the twist angle.
Inset: Unit cell (black rhomboid) for 19.4\,$^\circ$  twist angle contains 984 atoms. 
}
\label{f4}
\end{figure}

We now focus on an application of the $G_0W_0$ algorithm to transition-metal dichalcogenide heterobilayers which recently gained increased attention due to twist-angle dependent moir\'e potentials and interlayer excitons \cite{Zhang2020:NC,Gobato2022:NL,Huang2022:NN,Barre2022:S, Shabani2021,Naik2022,Alexeev2019Nature,Tang2020NatNano,Ma2021NL}.
Recent large-scale plane-wave-based \textit{GW} calculations on twisted heterostructures were limited to 75 atoms in the unit cell~\cite{Kundu2022}.
This $GW$ computation~\cite{Kundu2022}  has been described to be highly cumbersome and it was only achieved owing to an advanced accelerated large-scale version of the BerkeleyGW code which scales to entire leadership high-performance computers with more than half a million cores~\cite{DelBen2019,DelBen2020}.
Small unit cells with 75 atoms only allow for the study of heterobilayers with selected, large twist angles and absent atomic reconstruction. 
In order to illustrate the large-scale capabilities of our $G_0W_0$ algorithm beyond monolayers, we focus on the prototypical MoSe$_2$/WS$_2$ twisted heterostructures. On one hand, the different lattice parameters of MoSe$_2$ and WS$_2$ gives rise to a considerably large moiré periodicity at zero twist angle ($\sim8$ nm), thus requiring a large number of atoms in the structure. On the other hand, low-angle MoSe$_2$/WS$_2$ have shown an interesting interplay of intra- and inter-layer exciton hybridization because of the nearly degenerate conduction bands. This feature, however, is still under debate in the literature~\cite{Alexeev2019Nature,Tang2020NatNano,Zhang2020:NC,Ma2021NL,Gobato2022:NL}. The underlying electronic structure is thus crucial to resolve this controversy and is exactly the kind of problem that require large-scale $GW$ calculations. Here we considered MoSe$_2$/WS$_2$ moir\'e superstructures with twist angles between 9.3\,$^\circ$ and 26.6\,$^\circ$ (Fig.~\ref{f4}) that have corresponding unit cells of up to 984 atoms.
We emphasize that in all cases, the  strain  of the individual monolayers is ${<}\,0.01$\% compared to the experimentally determined lattice constants \cite{Schutte1987:JSSC, James1963:AC}, which is important because the bandgap is very sensitive to strain~\cite{Zollner2019:strain,Chaves2020}. 
The $G_0W_0$ bandgap of the MoSe$_2$/WS$_2$ bilayer depends on the twist angle changing from 1.86 eV (9.3\,$^\circ$) to 1.92 eV (26.8\,$^\circ$), in line with experimental observations of the exciton emission energy~\cite{Alexeev2019Nature}.
Our $GW$ calculation on the 984-atom heterostructure takes 42 hours on only 1536 cores which is a factor 30.000 faster than with a plane-wave algorithm, see estimate in the SI.
Such large-scale $GW$ calculations are an ideal starting point for further analyzing the electronic structure of these materials. 
For example, with our $GW$ algorithm, the calculation of deep moir\'e potentials~\cite{Shabani2021} are within reach, which are caused by atomic reconstruction and height variations. 
Both crucially influence the interlayer screening that is captured by the $GW$ method. 
On top of a $GW$ calculation, the Bethe-Salpeter equation~\cite{Onida2002,Blase2020} will enable the study of excitons in large-scale moir\'e structures.
Our computationally efficient scheme also holds great promise for nanoscale excited-state dynamics in low-dimensional materials.
Current state-of-the-art studies only report the dynamics in clean monolayers~\cite{Jiang2021,Chan2021,Perfetto2022} and models~\cite{Schluenzen2020,Tuovinen2023}.

Summarizing, we have presented a low-scaling $GW$ algorithm with periodic boundary conditions employing localized basis functions and the minimum image convention. 
The $GW$ algorithm is numerically precise and requires up to five orders of magnitude less floating point operations compared to plane-wave codes.
We carried out a $G_0W_0$ calculation on a MoSe$_2$/WS$_2$ heterostructure with 984 atoms in the unit cell which is an order of magnitude more than the state of the art~\cite{Kundu2022}. 
We are fully convinced that our \textit{GW} algorithm will enable routine applications of $GW$ and its time-dependent variants to low-dimensional, nanostructured materials that were previously computationally highly challenging.

\section*{Code and data availability}
The low-scaling \textit{GW} algorithm is implemented in the open-source CP2K package~\cite{Kuehne2020} which is freely available from github~\cite{CP2K}. 
Inputs and outputs of the calculations are also available on github~\cite{Inputs}.

\section*{Acknowledgment}

We thank Mauro Del Ben, Maria Camarasa-Gomez, Ferdinand Evers, Dorothea Golze, J\"urg Hutter, Ole Sch\"utt and Shridhar Shanbhag for helpful discussions. 
K.Z. and P.E.F.J.~acknowledge funding by the Deutsche Forschungsgemeinschaft (DFG, German Research Foundation) SFB 1277 (Project No.~314695032, projects B07 and B11), SPP 2244 (Project No.~443416183), and the European Union Horizon 2020 Research and Innovation Program under contract number 881603 (Graphene Flagship).
D. H.-P. acknowledges support from a Minerva Foundation grant 7135421, ERC Starting grant 101041159, and DFG through the SFB 1277 (No.~314695032, project B10).
J.W.~acknowledges funding by the DFG via the Emmy Noether
Programme (Project No.~503985532). 
The Gauss Centre for Supercomputing is acknowledged for providing computational resources on SuperMUC-NG at the Leibniz Supercomputing Centre under the project ID pn72pa. 
The \textsc{Quantum Espresso} and BerkeleyGW computations were carried out in the Max Planck Computing and Data Facility cluster.

\bibliography{maintext}
\bibliographystyle{apsrev4-2.bst}

\end{document}